\documentclass[conference]{IEEEtran}
\usepackage{graphicx, times, amsmath, amsfonts, amsthm, comment}
\usepackage{amssymb,epstopdf}
\usepackage{algorithmic}
\usepackage{algorithm}
\usepackage{color, soul}
\usepackage{mathtools}

\newcommand{\beq}{\begin{equation}}
\newcommand{\eeq}{\end{equation}}
\newcommand{\tbf}{\textbf}

\newcommand{\ud}{\mathrm{d}}

\newcommand*{\mathcolor}{}
\def\mathcolor#1#{\mathcoloraux{#1}}
\newcommand*{\mathcoloraux}[3]{%
  \protect\leavevmode
  \begingroup
    \color#1{#2}#3%
  \endgroup
}

\theoremstyle{plain}
\newtheorem{propcounter}{Proposition}
\newtheorem{proposition}[propcounter]{Proposition}
\theoremstyle{plain}

\theoremstyle{plain}

\theoremstyle{plain}
\newtheorem{assumecounter}{Assumption}
\newtheorem{assumption}[assumecounter]{Assumption}

\newcommand {\Ebb}{\mathbb{E}}

\newcommand {\Acal}{\mathcal{A}}
\newcommand {\Fcal}{\mathcal{F}}

\newcommand {\Kcal}{\mathcal{K}}
\newcommand {\Lcal}{\mathcal{L}}

\newcommand {\Ncal}{\mathcal{N}}

\begin{document}

\title{On Spectrum Sharing Among Micro-Operators in 5G}

\author{
\IEEEauthorblockN{Tachporn Sanguanpuak\IEEEauthorrefmark{1}, Sudarshan Guruacharya\IEEEauthorrefmark{2}, Ekram Hossain\IEEEauthorrefmark{2}, Nandana Rajatheva\IEEEauthorrefmark{1}, Matti Latva-aho\IEEEauthorrefmark{1}}
\IEEEauthorblockA{\IEEEauthorrefmark{1}Dept. of Commun. Eng., Univ. of Oulu, Finland; \IEEEauthorrefmark{2}Dept. Elec. \& Comp. Eng., Univ. of Manitoba, Canada.}
\IEEEauthorblockA{Email: \{tsanguan, rrajathe, matla\}@ee.oulu.fi; \{Sudarshan.Guruacharya, Ekram.Hossain\}@umanitoba.ca}
}\maketitle

\begin{abstract}

The growing demand in indoor small cell networks has given rise to the concept of micro-operators (MOs) for local service delivery.
We model and analyze a spectrum sharing system involving such MOs where a buyer MO buys multiple licensed subbands provided by the regulator. Also, all small cell base stations (SBSs) owned by a buyer MO can utilize multiple licensed subbands at the same time which are also used by other MOs. A deterministic model in which the location of the SBSs are known can lead to unwieldy problem formulation, when the number of SBSs is large. Subsequently, we adopt a stochastic geometric model of the SBS deployment instead of a deterministic model. Assuming that the locations of the SBSs can be modeled as a homogeneous Poisson point process, we find the downlink signal-to-interference-plus-noise ratio (SINR) coverage probability and average data rate for a typical user (UE) served by the buyer MO in a spectrum sharing environment. In order to satisfy the QoS constraint, we provide a greedy algorithm to find how many licensed subbands and which subband for the buyer MO to purchase from the regulator. We also derive the coverage probability of the buyer MO for interference the limited system.

\end{abstract}

\begin{IEEEkeywords}
Micro-operator, spectrum sharing, stochastic geometry, coverage probability, average data rate.
\end{IEEEkeywords}

\section{Introduction} \label{section:introduction}

In recent years, the concept of network infrastructure and spectrum sharing has been investigated to address the resource sharing problem for the network operators. On one hand, with the increasing of demand for mobile services, the under utilization of licensed spectrum auctioned off to the mobile network operators has become a bottleneck for the future growth of the industry \cite{Cisco2014}. On the other hand, in rural areas, where the demand can be low, the high cost of network infrastructure forces the network operators to charge high prices to their customers, this makes the service unaffordable to most people \cite{ITU2016}. One of the key aspects of the fifth generation (5G) mobile communication networks is to maximize the usage of existing network resources in terms of spectrum, infrastructure, and power while simultaneously minimizing the cost of purchasing resource, and reducing the energy consumption of the mobile devices \cite{ITU2016,Matinmikko2014}.


Nowadays wireless mobile service is given by typical mobile network operators which we refer to here as MNOs whose business model is to offer services with very high infrastructure investments and long investment period \cite{Matinmikko2014}. Regarding the research works based on the MNOs' points of view, in \cite{5GAmerica}, the concept of neutral host network deployment was proposed where the MNOs deploy cells in the best positions with optimal tuning to satisfy the quality-of-experience (QoE). The authors  also considered the sharing of other resources such as spectrum, rate, power adaptation, edge caching, and load balancing, which can be done across different virtual MNOs. In order to facilitate the local licensing models and to construct in high frequency band, the new innovations for mobile edge computing, network slicing, software defined networking, massive MIMO and wireless backhauling was proposed in \cite{Dehos2014}. In \cite{Jorswieck2014}, hardware demonstration of the benefit of inter-operator spectrum sharing was demonstrated. Resource sharing in the context of heterogeneous network and cloud RAN concepts was proposed \cite{Marcelo2015}.

Regarding stochastic geometry modelling of cellular systems owned by the MNOs, in \cite{Kibilda2016}, the point processes that model the spatial characteristics of the base stations (BSs) belonging to multiple MNOs was empirically studied, using the data from field surveys in Ireland, Poland, and UK. The authors conclude that the log-Gaussian Cox process is the best fit for the deployment patterns of the BSs. In \cite{Sanguanpuak2017}, the authors considered a single buyer--multiple seller BS infrastructure market as a Cournot oligopoly market. They modeled the locations of the base stations as a homogeneous Poisson point process and obtained the downlink signal-to-interference-plus-noise ratio (SINR) coverage probability for a user served by the buyer MNO in an infrastructure sharing environment. However, since the high volume of traffic densities comes from indoor environment such as hospitals, campuses, shopping malls, sport arenas, it leads to the problem that the traditional macro cellular networks become insufficient when the building penetration losses limit the indoor connectivity \cite{Matinmikko2017}. Hence, in the future, the business model, which is dominated by the MNOs, will become inadequate and various services cannot develop unless the wireless systems can response rapidly to the specific local traffic requirements.

One possible paradigm to address the above issue is to use the concept of micro operator (MO) to serve the specific local connectivity as firstly started in \cite{Ahokangas2016}. The authors identified the business model for the new MO concept. In \cite{Matinmikko2017}, the MO concept with the relation between MO to other stakeholders was proposed. Also the new spectrum regulation for MO network was provided. In this paper, we consider the scenario where one MO buys multiple licensed subbands from the regulator. In the spectrum sharing deployment, all the SBSs of the buyer MO can utilize multiple subbands. Also, the buyer MO allows the other MO who has low activity of UEs to utilize each subband at the same time. As such, for downlink transmission, each typical UE of the buyer MO experiences interference from the SBSs belonging to the other MO who is occupying that particular subband. We use results from stochastic geometric analysis of large-scale cellular networks to evaluate SINR outage probability and the average data rate for such a spectrum sharing system. In order to satisfy the QoS constraint in terms of coverage and the minimum required rate, we provide a greedy algorithm to find how many licensed subbands and which subbands the buyer MO has to purchase from the regulator. Then, in the simulation results, we show that spectrum sharing for MO network is beneficial for both coverage and average data rate.

%
%

\section{System Model, Assumptions, and QoS} \label{section:systemmodel}

\subsection{System Model and Assumptions}

Consider a system with the set of licensed spectrum subband $\Lcal = \{L_1,\ldots,L_2,\ldots, L_J\}$ owned by the regulator. We consider a system with $K+1$ micro-operators (MOs) given by the set $\Kcal = \{0,1,\ldots,K\}$, where each MO serves different local area such as, university, hospital and supermarket. Let MO-$0$ denotes the buyer who wants to buy the multiple licensed subbands from the regulator and MO-$k$, where $k \in \Kcal \backslash \{0\}$ is the other MO who is occupying the subband $L_j$, where $L_j \in \Lcal$. We assume that each MO-$k$, where $k \in \Kcal \backslash \{0\}$, has low level of UEs' activity. Let the set of small cell base stations (SBSs) owned by the MO-$k$ be given by $\Fcal_k$, where $k \in \Kcal$. Each of the SBSs and UEs is assumed to be equipped with a single antenna. The maximum transmit power of each SBS is $p_{\max}$ also a UE subscribing to an MO associates to the nearest SBS. The SBSs owned by different MOs are spatially distributed according to homogeneous Poisson point processes (PPPs). Let the spatial intensity of SBSs per unit area of MO-$k$ be denoted by $\lambda_k$, where $k \in \Kcal$. We can denote the overall net BSs intensity of all MOs as sum of all $\lambda_k$ by $\lambda_0 + \sum_{k\in \Kcal \backslash \{0\}} \lambda_k$.

The licensed subbands are orthogonal and hence there is no overlap between any two licensed subbands. We study when the buyer MO-$0$ buys multiple spectrum from the regulator and it allows other MO-$k$, where $k \in \Kcal \backslash \{0\}$, which has a low level of UEs' activity, to use each subband. For spectrum sharing among multiple MOs, we assume that the following assumptions hold:

\begin{assumption}
The MO-$0$ serves each typical UE of MO-$0$ itself using its own infrastructure while buying the licensed spectrum from the regulator. The typical UE of MO-$0$ associates with the nearest SBS in the set $\Fcal_0$ owned by the MO-$0$. Since each SBS can utilize multiple subbands, this implies that in each subband $L_j \in \Lcal$, the net SBS intensity that a typical UE of MO-$0$ can associate itself with is
\beq
\lambda_A = \lambda_0(L_j).
\label{eqn:lambdaA}
\eeq
\end{assumption}

\begin{assumption}
When the SBSs of MO-$0$ use the subband $L_j \in \Lcal$, MO-$0$ allows at most one $k$th MO, where $k \in \Kcal \backslash \{0\}$, to use the same subband simultaneously.
Also, each SBS of MO-$0$ is assumed to use multiple subbands. As for the downlink transmission from one SBS to each typical UE, each UE will receive transmissions from multiple subbands (channels) at the same time. In each subband $L_j$, the typical UE of MO-$0$ will experience interference from SBSs $\Fcal_0 \backslash \{0\}$ of MO-$0$ and $\Fcal_k$ of MO-$k$, where $k \in \Kcal\backslash\{0\}$.
When all the SBSs of MO-$0$ use the licensed subband $L_j \in \Lcal$, we have the intensity of interfering SBSs in each subband $L_j$ as
\beq
\lambda_I(L_j) = \lambda_0(L_j) + \nu_k \lambda_k(L_j), \quad \mbox{for} \quad k\neq 0.
\label{eqn:lambdaI_assump2}
\eeq
Here $0 \leq \nu_k \leq 1$ denotes the level of UEs' activity of MO-$k$ in the subband $L_k$.
\end{assumption}

\begin{figure}[h]
\centering
\includegraphics[height=3.2in, width=3.2 in, keepaspectratio = true]{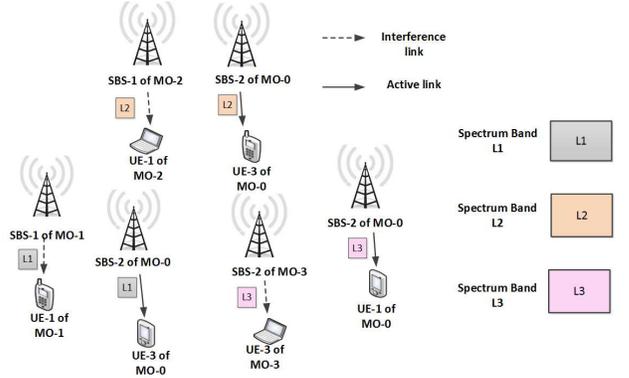}
\caption{The MO-$0$ buys only spectrum while using its own infrastructure.}
\label{fig:buy_spec}
\end{figure}

Fig. ~\ref{fig:buy_spec} illustrates the scenario when the MO-$0$ buys spectrum from the regulator. At the same time, there is another MO-$k$, where $k \in \Kcal \backslash \{0\}$, which uses the same subband. In each subband $L_j$, where $L_j \in \Lcal$, MO-$0$ allows at most one MO-$k$ to utilize the same subband. In this figure, MO-$0$ buys subbands $L_1$ and $L_3$ to serve its UE-$3$ which is associated with SBS-$2$. Since the SBSs of MO-$0$ can use multiple subbands (according to \tbf{Assumption 2}), in this figure, we assume that SBS-$2$ of MO-$0$ uses all three subbands $L_1$, $L_2$, and $L_3$ to serve its UEs.

We see that in the subband $L_1$, the transmit signal from the SBS-$1$ of MO-$1$ creates interference to the typical UE-$3$ of MO-$0$. Similarly, in subbands $L_2$ and $L_3$, the transmit signal from the SBS-$1$ of MO-$2$ and the transmit signal from the SBS-$2$ of MO-$3$ cause the interference to the typical UE-$3$ and the typical UE-$1$ of the MO-$0$, respectively.

\subsection{SINR Coverage and Average Rate}

Without loss of generality, we assume a typical UE of MO-$0$ located at the origin and associates with the nearest SBS of MO-$0$ from the set given by $\Fcal_0$. For the typical UE of MO-$0$, we will denote the nearest SBS from $\Fcal_0$ as SBS-$0$.

We assume that the message signal undergoes Rayleigh fading with the channel power gain given by $g_0$. Let $\alpha > 2$ denote the path-loss exponent for the path-loss model $r_0^{-\alpha}$, where $r_0$ is the distance between the typical UE and SBS-$0$. Let $\sigma^2$ denote the noise variance, and $p$ denote the transmit power of all the SBSs in MO-$0$, including SBS-$0$. The downlink ${\rm SINR}$ at the typical UE of MO-$0$ is
\beq
{\rm SINR} = \frac{g_0 r_0^{-\alpha}p}{I + \sigma^2},
\label{eqn:SINR}
\eeq
where $I$ is the interference experienced by a typical UE from the SBSs that operate on the spectrum $L_j$ where $L_j \in \Lcal$. These are the SBSs that belong to MO-$k$, where $k \in \Kcal \backslash \{0\}$ and MO-$0$. Thus, the interference $I = \sum_{i \in \Fcal_k \cup \Fcal_0 \backslash \{0\}}\psi_{i,j} g_i r_i^{-\alpha}p$. Here $g_i$ is the co-channel gain between the typical UE and interfering SBS-$i$, and $r_i$ is the distance between the typical UE and the interfering SBS-$i$, where $i \in \Fcal_k \cup \Fcal_0 \backslash \{0\}$. The transmit power of each SBS is $0 < p \leq p_{\max}$. Then, we assume $\psi_{i,j} \in \{0,1\}$ as a binary variable indicating whether the SBS-$i$ is active (if $\psi_{i,j} = 1$) or inactive (if $\psi_{i,j} = 0$) in spectrum subband $L_j$.

For a given threshold $T$, if ${\rm SINR} < T$ the UE is said to experience an outage (i.e., outage probability $P_{outage} (T) = \Pr ({\rm SINR} < T$). Likewise, if ${\rm SINR} > T$, then the UE is said to have coverage (i.e., coverage probability,  $P_c (T) = 1 - P_{outage} (T) = \Pr ({\rm SINR} \geq T$).  Given the SINR coverage probability, the average downlink transmission rate for a typical UE can be computed as
\begin{equation}
\Ebb[R] = \int_0^{\infty}P_c(e^T -1) \; dT.
\label{eqn:avgrate}
\end{equation}
We consider both the SINR coverage  probability  and a minimum average rate as the QoS metrics for a typical user.

\section{Analysis of SINR Coverage Probability}
\label{sec:avgcov_spec}

\subsection{SINR Coverage Probability When MO-0 Uses a Single Band}
Following to \cite[Theorem 1]{Andrews2011}, conditioning on the nearest BS at the distance $r$ from a typical UE, the coverage probability averaged over the plane is
\beq
P_c = \int_{r >0}\mathrm{Pr}({\rm SINR} > T \mid r)f_r(r) \; dr,
\eeq
where the probability density function (PDF) of $r$ can be obtained as \cite{Andrews2011}, $f_r(r) = e^{-\pi \lambda r^2} 2 \pi \lambda r dr$. Using the fact that the distribution of the channel gain follows an exponential distribution, a formula for a coverage probability of the typical UE when the BSs are distributed according to a homogeneous PPP of intensity $\lambda$ is derived in \cite[Eqn.2]{Andrews2011}. By observation, we can express the coverage probability in the most general form in terms of three components which are noise, interference and user association while each BS employs a constant power $p = 1/\mu$ as follows:
\beq
P_c = \int_{z > 0} \underbrace{e^{- \mu T z^{\alpha/2} \sigma^2}}_{\text{noise}} \underbrace{e^{-\pi(\lambda_I(\beta -1))z}}_{\text{interference}} \underbrace{e^{- \lambda_A \pi z} \pi \lambda_A}_{\text{user association}} dz,
\label{eqn:coverage_general}
\eeq
where $\lambda_I$ is the BS intensity causes interference to a typical UE, the UE associates with the closest BS (where the BS intensity is $\lambda_A$), the path-loss exponent is denoted as $\alpha$, and $\beta$ is given by
\beq
\beta = \frac{2 (T/p)^{2/\alpha}}{\alpha} \mathbb{E}_{g}[g^{2/\alpha} (\Gamma(-2/\alpha, T g/p))- \Gamma(-2/ \alpha)].
\label{eqn:beta}
\eeq

In particular, the general expression of the coverage probability in (\ref{eqn:coverage_general}) can be expressed as \cite[Theorem 1]{Andrews2011}
\beq
P_c = \pi \lambda_A \int_0^{\infty}\exp \{-(Az + Bz^{\alpha /2})\} dz,
\label{eqn:coverage-integral}
\eeq
where $A = \pi[(\lambda_I(\beta -1))+ \lambda_A]$ and $B = \frac{T \sigma^2}{p}$. When the interfering links undergo Rayleigh fading, $\beta = 1 + \rho(T,\alpha)$, where
\beq
\rho(T,\alpha) = T^{2/\alpha} \int_{T^{-2/\alpha}}^\infty (1+u^{\alpha/2})^{-1} \ud u.
\label{eqn:rho}
\eeq
For this special case, we see that $\beta$ is independent of transmit power. Except for $\alpha = 4$, the integral for $P_c$ cannot be evaluated in closed form. Nevertheless, a simple closed-form approximation for the general case, where $\alpha > 2$, and where both noise and intra-operator interference are present, can be given as \cite[Eqn.4]{Sudarshan2016}
\begin{align}
P_c \simeq \pi \lambda_A \left[ A+ \frac{\alpha}{2} \frac{B^{2/\alpha}}{\Gamma\big(\frac{2}{\alpha}\big)} \right]^{-1},
\label{eqn:coverage-approx}
\end{align}
in which $\Gamma(z,a) = \int_z^{\infty} x^{a-1}e^{-x} dx$ is the upper incomplete Gamma function.


\subsection{SINR Coverage Probability Under Spectrum Sharing}

In our spectrum sharing model, the regulator sells the licensed subband to the MO-$0$ while some of the SBSs of the MO-$k$, where $k \in \Kcal \backslash \{0\}$, are using the same subband. Also, we consider that all SBSs of MO-$0$ are using $L$ licensed subbands (where $\mid \Lcal \mid = L$, $\mid \Acal \mid$ denotes the cardinality of a set $\Acal$) at the same time. Due to the fact that MO-$0$ buys only spectrum, the UEs of MO-$0$ always associates to the SBS-$0$, where $\{0\} \in \Fcal_0$. For our system, since the SBSs of MO-$0$ utilize multiple subbands at the same time, we have to modify the formulas (\ref{eqn:coverage_general}) and (\ref{eqn:coverage-integral}) and show that a more general coverage formula is given as follows:

\begin{proposition}
Under \tbf{Assumption 1} and \tbf{Assumption 2}, the coverage probability of a typical UE of MO-$0$ is
\beq
P_c = \sum_{L_j \in \Lcal}P_c(L_j) \Pr(L_j),
\label{eqn:Pc_tot}
\eeq
where $P_c(L_j)$ denotes the coverage probability of the UE of MO-$0$ using band $L_j$ and $Pr(L_j)$ is the probability of the typical UE of MO-$0$ using band $L_j$.
\end{proposition}
By using the conditional probability, we obtain the coverage probability $P_c(L_j)$ of the MO-$0$ when it uses the band $L_j$ given the probability of using the band $L_j$, $\Pr(L_j)$, in which, $\Pr(L_j) \in \{0,1\}$ indicating whether the MO-$0$ buys and uses the licensed band $L_j$ (if $\Pr(L_j) =1$) or the MO-$0$ does not use the subband $L_j$ (if $\Pr(L_j) =0$). Then, we take the summation over the bands $L$, where $\mid \Lcal \mid = L$.

Let us consider he case when $\lambda_I(L_j) = \lambda_0(L_j) + \nu_k \lambda_k(L_j)$ and $\nu_k = 1$.
Since all SBSs $\Fcal_0$ of MO-$0$ using the licensed band and in each aubband $L_k$, while there is MO-$k$ who has low activity UEs occupying that band. We can denote $\nu_k = 1$ in (\ref{eqn:lambdaI_assump2}). The intensity of interfering SBSs in the band $L_k$ is  $\lambda_I(L_k) = \lambda_0(L_k) + \lambda_k(L_k)$, where $k \in \Kcal \backslash \{0\}$.

\begin{proposition}
Under \tbf{Assumption 1} and \tbf{Assumption 2}, the coverage probability of a typical UE of MNO-$0$ using the band $L_j$, where $L_j \in \Lcal$,  is given by
\beq
P_c(L_j) = \pi \lambda_A \int_0^{\infty} \exp\{-(A_1 z + B z^{\alpha/2})\} dz,
\label{eqn:Pc_Lk_nu1}
\eeq
where $A_1 = \pi ((\lambda_0(L_k) + \lambda_k(L_k))\beta - \lambda_k(L_k))$, and by \tbf{Assumption 1}, we can assume $\lambda_A = \lambda_0$. Also,  $\beta$ and $B$ are given by (\ref{eqn:beta}) and (\ref{eqn:coverage-integral}), respectively. Then, we can approximate $P_c(L_j)$ in (\ref{eqn:Pc_Lk_nu1}) using (\ref{eqn:coverage-approx}) as
\beq
P_c(L_j) = \frac{\pi \lambda_0(L_j)}{A_1 + \frac{\alpha}{2}\frac{B^{2/\alpha}}{\Gamma\left(\frac{2}{\alpha}\right)}}.
\eeq

Without loss of generality, we can assume $\Pr(L_j) =1$, where $L_j \in \Lcal$. Hence, we obtain $P_c$ as
\beq
P_c =  \sum_{L_j \in \Lcal} \frac{ \pi \lambda_0(L_j)}{ A_1 +  \frac{\alpha}{2} \frac{B^{2/\alpha}}{\Gamma\big(\frac{2}{\alpha}\big)}}.
\label{eqn:Pctot_nu1}
\eeq
\end{proposition}
\begin{IEEEproof}
We obtain $P_c(L_j)$ in (\ref{eqn:Pc_Lk_nu1}) by using an expression in (\ref{eqn:coverage-integral}) after that evaluating (\ref{eqn:Pc_Lk_nu1}) by using a closed form approximation in (\ref{eqn:coverage-approx}), where $B$ and $\beta$ are the same as (\ref{eqn:coverage-integral}). Then, we can express $P_c$ by substituting $P_c(L_j)$ in (\ref{eqn:Pc_tot}) while assuming $\Pr(L_j) = 1$.
\end{IEEEproof}

Next, we consider the scenario when the system becomes \emph{``interference-limited''}, which occurs when $\sigma^2 \to 0$.
\begin{proposition}
The coverage probability for interference-limited case when the MO-$0$ using the subband $L_j$, where $L_j \in \Lcal$, can be expressed as
\beq
P_c = \sum_{L_j \in \Lcal} \frac{1}{\beta + (\beta-1)\frac{\lambda_k(L_j)}{\lambda_0(L_j)}},
\eeq
in which, the $\lambda_k(L_j)$ and $\lambda_0(L_j)$ are the SBS intensity of MO-$k$ and MO-$0$ using the band $L_j$, respectively.
\end{proposition}
\begin{IEEEproof}
Let $L = |\Lcal|$. For interference-limited case, i.e., $B \to 0$ in (\ref{eqn:Pctot_nu1}), and let $C=\frac{\alpha}{2} \frac{B^{2/\alpha}}{\Gamma\big(\frac{2}{\alpha}\big)}$ as such we can neglect $C$ when the system becomes interference limited. After simplifying (\ref{eqn:Pctot_nu1}) while assuming $C=0$, we have the required result.
\end{IEEEproof}

%

\section{Amount of Spectrum Bands Required to Satisfy the QoS}
\label{sec:avgrate}

The expected rate can be derived using the closed form approximation of coverage probability from (\ref{eqn:coverage-approx}) with the assumption that interference is Rayleigh fading where $\beta = 1 + \rho(\hat{T},\alpha)$, given in (\ref{eqn:rho}), with $\hat{T} = e^T - 1$. The expected rate $\Ebb[R]$ for the general case can be given by
\begin{align}
\Ebb[R] &= \pi \lambda_{A} \int_0^{\infty}\bigg[ \pi \big( \lambda_{I} \hat{T}^{2/\alpha} \int_{\hat{T}^{-2/\alpha}}^{\infty} (1+ u^{\alpha/2})^{-1}du + \lambda_A \big) \nonumber \\
& \qquad + \frac{\alpha}{2}\frac{B^{2/\alpha}}{\Gamma(2/\alpha)}\bigg]^{-1} d\hat{T}, \nonumber \\
&= \pi \lambda_{A} \int_0^{\infty}\bigg[ \pi \big( \lambda_{I} \frac{2 \hat{T}}{\alpha -2} {}_{2}F_{1}\big(1, 1-\frac{2}{\alpha};2-\frac{2}{\alpha};1-\hat{T}\big) \nonumber \\
& \qquad + \lambda_A \big) + \frac{\alpha}{2}\frac{B^{2/\alpha}}{\Gamma(2/\alpha)}\bigg]^{-1} d\hat{T},
\label{eqn:rateapprox}
\end{align}
where $\Gamma(z)$, and ${}_{2}F_{1}(a,b,c,z)$ are the Gamma function, and the  Hypergeometric function, respectively. The average rate in (\ref{eqn:rateapprox}) is valid for any real values of $\alpha > 2$, $T >0$ and can be evaluated by numerical integration techniques.

\begin{proposition}
The expected rate of a typical UE of MO-$0$ using multiple subbands $L$, while in each subband $L_j \in \Lcal$, there is other MO-$k$ who has low activity UE using the same subband is
\begin{align}
&\Ebb[R] = \sum_{L_j \in \Lcal}\pi \lambda_{0}(L_j) \int_0^{\infty}\bigg[ \pi \big\{ (\lambda_{0}(L_j)+\lambda_k(L_j)) \left(\frac{2 \hat{T}}{\alpha -2}\right) \nonumber\\
&\times {}_{2}F_{1}\big(1, 1-\frac{2}{\alpha};2-\frac{2}{\alpha};1-\hat{T}\big)+ \lambda_0 \big\} + \frac{\alpha}{2}\frac{B^{2/\alpha}}{\Gamma(2/\alpha)}\bigg]^{-1} d\hat{T}.
\label{eqn:rateapprox_Assump2}
\end{align}
\end{proposition}

\begin{IEEEproof}
As the average rate is obtained by taking an integration of $P_c$ with respect to $\hat{T}$ from zero to infinity, and we use $P_c$ from (\ref{eqn:Pctot_nu1}). Following \tbf{Assumption 1} and \tbf{Assumption 2}, when MO-$0$ uses the band $L_j$ to serve its UE, we substitute $\lambda_A = \lambda_0(L_j)$ and $\lambda_I = \lambda_0(L_j) + \lambda_k(L_j)$ in (\ref{eqn:rateapprox}). We obtain the result in (\ref{eqn:rateapprox_Assump2}).
\end{IEEEproof}

Let us further assume that the MO-$0$ wants to ensure that the coverage probability of a typical UE satisfies the QoS constraint
\beq
P_c \geq 1 - \epsilon,
\label{eqn:QoS-constrain}
\eeq
where $0< \epsilon < 1$ is some arbitrary value.

In order to satisfy the QoS constraint in (\ref{eqn:QoS-constrain}), the buyer MO-$0$ will select the number of licensed subbands needed, at minimum cost, such that it can serve its UEs guaranteeing some QoS. Due to the fact that, there is a cost associated with each subband denoted as $q_{L_k}$, then we use greedy algorithm to find  which subband and how many of them the MO-$0$ will buy to satisfy the QoS of its UE. Let $N = \sum_{l\in \Lcal} N_l$ denote the minimum number of licensed subbands needed for the MO-$0$. For the QoS condition of the minimum rate requirement needed at each UE of the buyer MO-$0$ denoted by $R_{\text{min}}$ to be feasible, the minimum number of licensed subbands needed must satisfy
\beq
N \times \Ebb[R] \geq R_{\text{min}},
\label{eqn:min_rate_required}
\eeq
Let $\Ebb[R]$ denote the expected rate at the UE of MO-$0$ obtained from (\ref{eqn:rateapprox}). For MO-$0$, the maximum number of bands  required in order to satisfy both SINR coverage  and the minimum rate required at it's typical UE is
\beq
L_{\max} = \text{max}\{N, M\},
\eeq
where $M = \sum_{l\in \Lcal}M_l$ is the number of licensed subbands needed to satisfy the rate constraint. Let us now propose a simple greedy algorithm in \cite[Chap 17.1]{Korte2012} to select for the MO-$0$ to select which licensed subband and how many licensed aubbands that the MO-$0$ will purchase from the regulator. The greedy algorithm is provided in \textbf{ Algorithm \ref{alg:GreedyAlgo}}. The idea behind this greedy algorithm is as follows: We first sort the licensed subbands $L_j \in \Lcal$ according to the cost per subband $q_{L_j}$ in an ascending order. After using the greedy algorithm, we obtain $L_{\max}$ number of licensed subbands in order to satisfy both of the coverage QoS and the minimum rate needed at the UE.

\begin{algorithm}
\caption{Greedy Algorithm}
\label{alg:GreedyAlgo}
 \begin{algorithmic}[1]
 \STATE Sort the subbands by $q_{L_j}$ in ascending order such that $q_{\pi_1} \leq q_{\pi_2} \cdots \leq q_{\pi_{L_J}}$
 \FOR{$i = 1$ \TO $L_J$}
      \STATE Set $\Ncal = \{\pi_1,\ldots, \pi_{L_j} \}$ where $|\Ncal| = N$
       \IF {$\sum_{l \in \Ncal} \frac{ \pi \lambda_0(l)}{ A_1 +  \frac{\alpha}{2} \frac{B^{2/\alpha}}{\Gamma\big(\frac{2}{\alpha}\big)}} \geq 1-\epsilon$}
       \STATE Compute $\sum_{l\in \Ncal}M_l = M$.
       \ENDIF
       \IF {$\sum_{l\in \Lcal} N_l \times \Ebb[R] \geq R_{\text{min}}$}
       \STATE Compute $\sum_{l\in \Ncal}N_l = N$.
       \STATE Terminate
      \ENDIF
      \STATE Compute $L_{\max} = \max\{N,M\}$.
 \ENDFOR
 \STATE Compute $P_c$ using (\ref{eqn:Pctot_nu1}) with $L = L_{\max}$.
 \STATE Compute $\Ebb[R]$ using (\ref{eqn:rateapprox}) with $L = L_{\max}$.
 \end{algorithmic}
 \end{algorithm}

\section{Numerical Results}\label{section:Numerical Results}
We assume that the SBSs are spatially distributed according to homogeneous PPP inside a circular area of $500$ meter radius for all $K+1$ MOs. The MOs are assumed to have the same intensity of SBSs per unit area. The maximum transmit power of each SBS is $p_{\max} = 10$ dBm. The path-loss exponent is $\alpha = 4$, and noise power $\sigma^2 = -150$ dBm. Each SBS from all MOs transmits at the maximum power. The coverage probability is obtained from (\ref{eqn:Pctot_nu1}) and the average data rate is plotted accordingly. We illustrate the simulation results for the case when the buyer MO-$0$ purchases multiple licensed subbands while assuming that the cost of each subband is equal.

\subsection{Effect of Changing the Average Number of SBSs of MO-$0$ per Unit Area}

\begin{figure}[h]
\centering
\includegraphics[height=3.3 in, width=3.4 in, keepaspectratio = true]{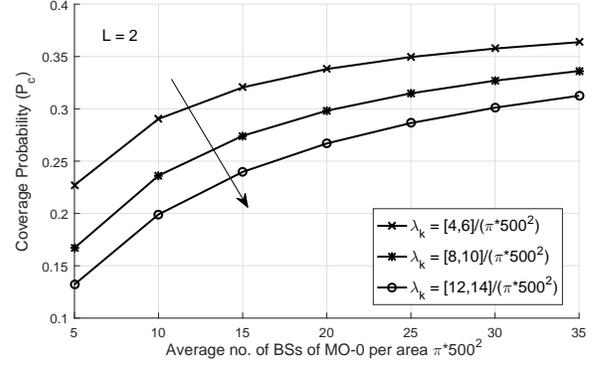}
\caption{The coverage probability of the typical UE of MO-$0$ while increasing the number of MO-$0$ per unit area.}
\label{fig:Pc_changeBS}
\end{figure}

\begin{figure}[h]
\centering
\includegraphics[height=3.3 in, width=3.4 in, keepaspectratio = true]{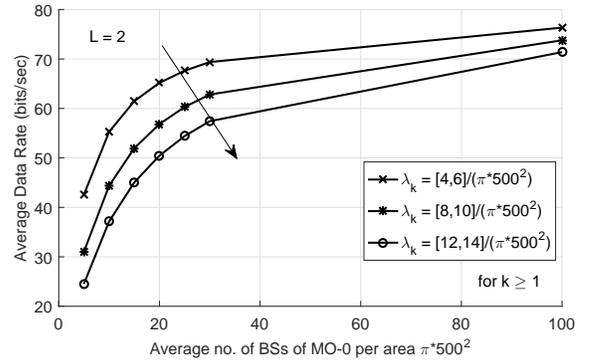}
\caption{The average data rate of the typical UE of MO-$0$ while increasing the number of MO-$0$ per unit area.}
\label{fig:AvgRate_changeBS}
\end{figure}

In Fig.~\ref{fig:Pc_changeBS} and Fig.~\ref{fig:AvgRate_changeBS}, the simulation parameters are as follows: the SINR threshold at each typical UE of MO-$0$ is set to $T = 10$ dB. We consider when the regulator sells two licensed subbands and each subband has one MO-$k$, where $k\in \Kcal \backslash \{0\}$ is occupying the subband. The MO-$0$ is using its own infrastructure to serve its UE. We consider the cases when the MO-$0$ buys two licensed subband, it means that each SBS of MO-$0$ utilizes two licensed spectrum at the same time. Fig.~\ref{fig:Pc_changeBS} plots the coverage probability when the average number of SBSs of MO-$0$ per unit area $\pi \times 500^2$ is increased. When average number of SBSs of MO-$0$ increases, the coverage probability of MO-$0$ is also increased. By increasing the SBSs intensity of the MO-$k$, where $k \in \Kcal \backslash \{0\}$ in each subband $L_k$, we see that the coverage probability of MO-$0$ decreases. This is not surprising since the SBS intensity of MO-$k$ in which $k \in \Kcal \backslash \{0\}$ causes interference in each band $L_k$.

In Fig.~\ref{fig:AvgRate_changeBS}, the average data rate of MO-$0$ is shown. When the average number of SBSs of MO-$0$ increases, the average data rate of MO-$0$ is increased. We also show when the average number of SBS of MO-$0$ is very high such that it tends to infinity, the average data rate tends to saturate at one value. Also when the SBS intensity of MO-$k$, where $k \in \Kcal \backslash \{0\}$ increases, the average data rate decreases.

\subsection{Effect of Changing the SINR Threshold}

\begin{figure}[h]
\centering
\includegraphics[height=3.3 in, width=3.4 in, keepaspectratio = true]{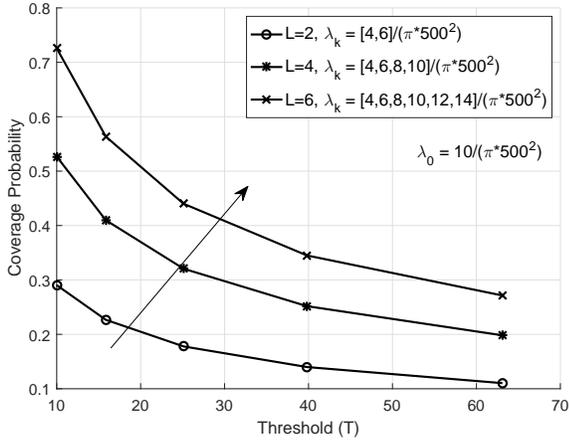}
\caption{The coverage probability of the typical UE of MO-$0$ when increasing the SINR threshold (T).}
\label{fig:Pc_T}
\end{figure}

\begin{figure}[h]
\centering
\includegraphics[height=3.3 in, width=3.4 in, keepaspectratio = true]{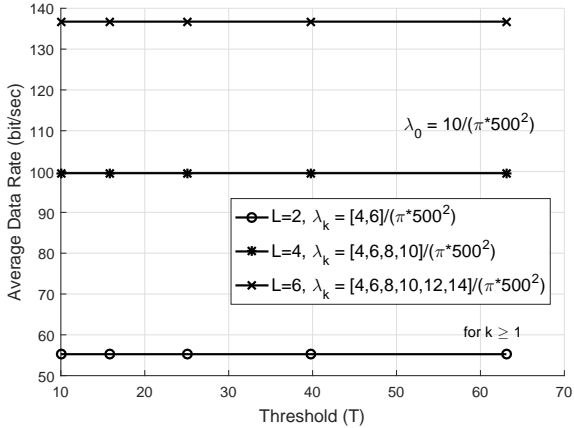}
\caption{The average data rate of the typical UE of MO-$0$ with increasing the SINR threshold ($T$) at the UE of MO-$0$.}
\label{fig:Ravg_T}
\end{figure}

In Fig.~\ref{fig:Pc_T} and Fig.~\ref{fig:Ravg_T}, we illustrate the coverage probability and the average rate of MO-$0$ when the increasing of SINR threshold ($T$) at each UE of MO-$0$. The SBS intensity of MO-$0$ is set to $\lambda_0 = 10/(\pi*500^2)$. We see that when $T$ increases, the SINR coverage probability of MO-$0$ decreases. We also consider when the MO-$0$ buys two, four and six licensed bands. In Fig.~\ref{fig:Pc_T}, we see that when the number of licensed subbands increases, the coverage probability  also increases. Although for the case of the MO-$0$ buys six licensed subbands with high SBSs intensity of MO-$k$, the coverage of MO-$0$ still increases. In Fig.~\ref{fig:Ravg_T}, we see that the average data rate remains constant when the SINR threshold increases. This is because, we first calculate the coverage probability for each threshold ($T$). Then, integrate the coverage probability $P_c$ with respect to the threshold, the expression of the average data rate becomes (\ref{eqn:rateapprox_Assump2}). As such, the average data rate remains constant with the changing of $T$. However, the average data rate increases significantly when MO-$0$ buys more bands.

\section{Conclusion}\label{section:Conclusion}
We have studied the problem of spectrum sharing among  multiple micro-operators (MOs) using stochastic geometry, where the buyer MO buys multiple subbands from a regulator. Also, the buyer MO allows other MOs to utilize the same subband. We have first analyzed the downlink coverage probability for a typical user served by a buyer MO, and subsequently, we have derived the average data rate. Both the SINR coverage and a minimum rate requirement are considered as the QoS metrics. In order to satisfy the QoS constraints of the typical user served by the buyer MO, we have provided a greedy algorithm to find how many subbands and which subbands for the buyer MO to purchase from the regulator. Both the coverage and the average data rate of the buyer MO  increase when the buyer MO buys more licensed subbands. However, when the average number of SBS per unit area of the buyer MO  increases and approaches infinity, the average data rate for a typical user served by the buyer  MO saturates to a single value.

\section*{Acknowledgments}
For this work, the authors would like to acknowledge the
support from Finnish Funding Agency for Technology and In-
novation (TEKES), Nokia Networks, Anite Telecoms, Broadcom Communications Finland, Elektrobit Wireless Communications and Infotech Oulu Graduate School.
Also, support from the Natural Sciences and Engineering Research Council of Canada (NSERC) is acknowledged.

\bibliographystyle{IEEE}

\begin{thebibliography}{1}

\bibitem{Cisco2014}
Cisco, ``Visual Networking Index: Global Mobile Data Traffic Forecast
Update,'' Cisco, San Jose, CA, USA, Feb. 2014, 2013--2018.


\bibitem{ITU2016}
International Telecommunications Union, ``Mobile infrastructure sharing,'' {\em ITU News Magazine}. [Online]. Accessed: 28 Oct. 2016. Available: {\em http://www.itu.int/itunews/manager/display.asp?lang=en
\&year=2008\&issue=02\&ipage=sharingInfrastructure-mobile}.


\bibitem{Matinmikko2014}
M.~Matinmikko {\em et al.}, ``Spectrum sharing using licensed shared access: the concept and its workflow for LTE-advanced networks,'' {\em IEEE Wireless Commun.} vol. 21, no. 2, pp. 72--79, 2014.

\bibitem{5GAmerica}
5G America, ``Multi-operator and neutral host small cells, Drivers, architectures, planning and regulation,'' 2016.

\bibitem{Dehos2014}
C.~Dehos, {\em et al.}, ``Millimeter-wave access and backhauling: the solution to the exponential data traffic increase in 5G mobile communications systems?.'' {\em IEEE Commun. Magazine}, vol. 52, pp. 88--95, 2014.


\bibitem{Jorswieck2014}
E.A.~Jorswieck {\em et al.}, ``Spectrum sharing improves the network efficiency for cellular operators,'' {\em IEEE Commun. Mag.}, vol. 52, no. 3, pp. 129--136, 2014.

\bibitem{Marcelo2015}
M.~Antonio Marotta, {\em et al.},``Resource sharing in heterogeneous cloud radio access networks,'' {\em IEEE Wireless Commun.}, vol. 22, pp. 74--82, 2015.


\bibitem{Kibilda2016}
J.~Kibiłda, B.~Galkin, and L. A.~DaSilva, ``Modelling multi-operator base station deployment patterns in cellular networks,'' {\em IEEE Trans. on Mobile Computing}, vol. 15, no. 12, pp. 3087--3099, Dec. 1 2016.


\bibitem{Sanguanpuak2017}
T.~Sanguanpuak {\em et al.}, ``Inter-operator infrastructure sharing: Trade-offs and Market,'' {\em Proc. IEEE Int. Conf. Commun. (ICC)}, to appear.




\bibitem{Matinmikko2017}
M.~Matinmikko {\em et al.}, ``Micro operaors to boost local service delivery in 5G,'' {\em  Wireless Personal Communications}, to appear.


\bibitem{Ahokangas2016}
P.~Ahokangas {\em et al.}, ``Future micro operators business models in 5G,'' {\em Proc. Int. Conf. on Restructuring of the Global Economy (ROGE)}, 2016.



\bibitem{Andrews2011}
J.G.~Andrews, F.~Baccelli, and R.K.~Ganti, ``A tractable approach to coverage and rate in cellular networks,'' {\em IEEE Trans. on Commun.}, vol. 59, no. 11, pp. 3122--3134, Nov. 2011.


\bibitem{Sudarshan2016}
S.~Guruacharya, H.~Tabassum, and E.~Hossain, ``Integral approximations for coverage probability,'' {\em IEEE Wireless Commun. Letters }, vol. 5, issue 1, pp. 24--27, Feb. 2016.


\bibitem{Korte2012}
B.~Korte and J.~Vygen, {\em Combinatorial Optimization: Theory and Algorithms}. 5th ed., Springer, 2012.


\end{thebibliography}

\end{document}